\documentclass[aps,prd,showpacs,floatfix]{revtex4}
\usepackage{amsmath,amsfonts}
\usepackage[dvips]{graphicx}

\newcommand{\ud}{\mathrm{d}}

\begin{document}

\title{Route to Lambda in conformally coupled phantom cosmology}
\author{Orest Hrycyna}
\email{hrycyna@kul.lublin.pl}
\affiliation{Department of Theoretical Physics, Faculty of Philosophy, 
The John Paul II Catholic University of Lublin, Al. Rac{\l}awickie 14, 
20-950 Lublin, Poland}
\affiliation{Astronomical Observatory, Jagiellonian University,
Orla 171, 30-244 Krak{\'o}w, Poland}
\author{Marek Szyd{\l}owski}
\email{uoszydlo@cyf-kr.edu.pl}
\affiliation{Astronomical Observatory, Jagiellonian University,
Orla 171, 30-244 Krak{\'o}w, Poland}
\affiliation{Mark Kac Complex Systems Research Centre, Jagiellonian University,
Reymonta 4, 30-059 Krak{\'o}w, Poland}


\begin{abstract}
In this letter we investigate acceleration in the flat cosmological model with 
a conformally coupled phantom field and we show that acceleration is its 
generic feature. We reduce the dynamics of the model to a $3$-dimensional 
dynamical system and analyze it on a invariant $2$-dimensional submanifold. 
Then the concordance FRW model with the cosmological constant $\Lambda$ is a 
global attractor situated on a $2$-dimensional invariant space. We also study 
the behaviour near this attractor, which can be approximated by the dynamics of
the linearized part of the system. We demonstrate that trajectories of the 
conformally coupled phantom scalar field with a simple quadratic potential
crosses the cosmological constant barrier infinitely many times in the phase
space. The universal behaviour of the scalar field and its potential is also
calculated. We conclude that the phantom scalar field conformally coupled to 
gravity gives a natural dynamical mechanism of concentration of the equation of 
state coefficient around the magical value $w_{\text{eff}}=-1$. We demonstrate 
route to Lambda through the infinite times crossing the $w_{\text{eff}}=-1$ 
phantom divide.
\end{abstract}


\maketitle

At present the scalar fields play a crucial role in modern cosmology. In an
inflationary scenario they generate an exponential rate of evolution of the 
universe as well as density fluctuations due to vacuum energy. The Lagrangian 
for a phantom scalar field on the background of the Friedmann-Robertson-Walker 
(FRW) universe is assumed in the form
\begin{equation}
\mathcal{L}_{\psi}=\frac{1}{2} [ g^{\mu \nu} \partial_{\mu}\psi
\partial_{\nu}\psi + \xi R \psi^{2} - 2 U(\psi)],
\label{eq:1}
\end{equation}
where $g^{\mu \nu}$ is the metric of the spacetime manifold, $\psi=\psi(t)$, 
$t$ is the cosmological time, $R=R(g)$ is the Ricci scalar for the spacetime 
metric$g$, $\xi$ is a coupling constant which assumes zero for a scalar field 
minimally coupled to gravity and $1/6$ for a conformally coupled scalar field, 
$U(\psi)$ is a potential of the scalar field.

The minimally coupled slowly evolving scalar fields with a potential function
$U(\psi)$ are good candidates for a description of dark energy. In this model,
called quintessence \cite{Wetterich:1988, Ratra:1988}, the energy density and 
pressure from the scalar field are $\rho_{\psi}=-1/2 \dot{\psi}^{2}+U(\psi)$, 
$p_{\psi}=-1/2\dot{\psi}^{2}-U(\psi)$. From recent studies of observational 
constraints we obtain that $w_{\psi} \equiv p_{\psi}/\rho_{\psi} < -0.55$ 
\cite{Chae:2002}. This model has been also extended to the case of a complex 
scalar field \cite{Gu:2001,Gao:2002}.

Observations of distant supernovae support the cosmological constant term which
corresponds to the case $\dot{\psi} \simeq 0$. Then we obtain that
$w_{\psi}=-1$. But there emerge two problems in this context. Namely, the fine
tuning and the cosmic coincidence problems. The first problem comes from the
quantum field theory where the vacuum expectation value is of $123$ orders of 
magnitude larger than the observed value of $10^{-47}\text{GeV}^{4}$. The lack 
of a fundamental mechanism which sets the cosmological constant almost zero is 
called the cosmological constant problem. The second problem called ``cosmic 
conundrum'' is a question why the energy densities of both dark energy and dark 
matter are nearly equal at the present epoch.

One of the solutions to this problem offers the idea of quintessence, which is 
a version of the time varying cosmological constant conception. Quintessence 
solves the first problem through the decaying $\Lambda$ term from the beginning 
of the Universe to a small value observed at the present epoch. Also the ratio 
of energy density of this field to the matter density increases slowly during 
the expansion of the Universe because the specific feature of this model is the 
variation of the coefficient of the equation of state with respect to time. The 
quintessence models \cite{Ratra:1988,Caldwell:1998} describe the dark energy 
with the time varying equation of state for which $w_{X}>-1$, but recently 
quintessence models have been extended to the phantom quintessence models with 
$w_{X}<-1$. In this class of models the weak energy condition is violated and 
such a theoretical possibility is realized by a scalar field with a switched 
sign in the kinetic term $\dot{\psi}^{2} \to -\dot{\psi}^{2}$ 
\cite{Caldwell:2002, Dabrowski:2003, Copeland:2006}. From theoretical point of 
view it is necessary to explore different evolutional scenarios for dark energy 
which provide a simple and natural transition to $w_{X}=-1$. The methods of 
dynamical systems with notion of attractor (a limit set with an open inset) 
offers the possibility of description of transition trajectories to the regime 
with $w_{X}=-1$. Moreover they demonstrate whether this mechanism is generic.

Inflation and quintessence with non-minimal coupling constant are studied in
the context of formulation of necessary conditions for the acceleration of the
universe \cite{Faraoni:2000wk} (see also \cite{Faraoni:2000nt, Bellini:2002zr}).
We can find two important arguments which favour the choice of conformal
coupling over $\xi \ne 1/6$.
The first, equation for the massless scalar field is conformally invariant
\cite{Birrell:1984ix, Penrose:1964ge}.
The second argument is that if the scalar field satisfy Klein-Gordon equation
in the curved space then $\psi$ does not violate the equivalence principle, and
$\xi$ is forced to assume the value $1/6$ \cite{Faraoni:1996rf}.

While recent astronomical observations give support that the equation of state 
parameter for dark energy is close to constant value $-1$ they do not give 
a ``corridor'' around this value. Moreover, Alam et al. \cite{Alam:2004} 
pointed out that evolving state parameter is favoured over constant $w_{X}=-1$. 
The first step in the direction of description of the dynamics of the dark 
energy seems to be investigation of the system with evolving dark energy in the 
close neighbourhood of the value $w_{X}=-1$. For this aim we linearize 
dynamical system at this critical point and then describe the system in a good 
approximation (following the Hartman-Grobman theorem \cite{Perko:1991}) by its 
linearized part. 

Other dark energy models like the Chaplygin gas model \cite{Kamenshchik:2001cp,
Bilic:2002vm, Bento:2002ps},
\cite[and references therein]{Copeland:2006} and the model with tachyonic 
matter can also be interpreted in terms of a scalar field with some form of 
a potential function.

Recent applications of the Bayesian framework \cite{Kurek:2007, Szydlowski:2006kk,
Szydlowski:2005g, Szydlowski:2006g1, Szydlowski:2006g2} of model selection to
the broad class of cosmological models with acceleration indicate that a 
posteriori probability for the $\Lambda$CDM model is $96\%$. Therefore the
explanation why the current universe is such close to the $\Lambda$CDM model 
seems to be a major challenge for modern theoretical cosmology. 

In this letter we present the simplest mechanism of concentration around 
$w_{X}=-1$ basing on the influence of
a single scalar field conformally coupled to gravity acting in the radiation
epoch. Phantom cosmology non-minimally coupled to the Ricci scalar was explored
in the context of superquintesence ($w_{X}<-1$) by Faraoni \cite{Faraoni:2002,
Faraoni:2005} and there was pointed out that the superacceleration regime can be
achieved by the conformally coupled scalar field in contrast to the minimally
coupled scalar field. 

Let us consider the flat FRW model which contains a negative kinetic scalar 
field conformally coupled to gravity $(\xi=1/6)$ (phantom) with the potential 
function $U(\psi)$. For the simplicity of presentation we assume $U(\psi) 
\propto \psi^{2}$. In this model the phantom scalar field is coupled to gravity 
via the term $\xi R \psi^{2}$. We consider massive scalar fields (for recent 
discussion of cosmological implications of massive and massless scalar fields 
see \cite{Jankiewicz:2006}). The dynamics of a non-minimally coupled scalar 
field for some self-interacting potential $U(\psi)$ and for an arbitrary $\xi$ 
is equivalent to the action of the phantom scalar field (which behaves like a 
perfect fluid) with energy density $\rho_{\psi}$ and pressure $p_{\psi}$ 
\cite{Gunzig:2000}
\begin{equation}
\rho_{\psi}= -\frac{1}{2}\dot{\psi}^{2} + U(\psi) - 3\xi H^{2} \psi^{2} - 3\xi
H(\psi^{2})\dot{},
\label{eq:2}
\end{equation}
\begin{equation}
p_{\psi}=-\frac{1}{2}\dot{\psi}^{2} - U(\psi) +\xi \Big[ 2H(\psi^{2})\dot{} +
(\psi^{2})\ddot{} \Big] + \xi \Big[ 2\dot{H}+3H^{2} \Big]\psi^{2},
\label{eq:3}
\end{equation}
where the conservation condition $\dot{\rho}_{\psi}=-3H(\rho_{\psi}+p_{\psi})$
gives rise to the equation of motion for the field
\begin{equation}
\ddot{\psi} + 3H\dot{\psi} + \xi R \psi^{2} - U'(\psi) = 0,
\label{eq:4}
\end{equation}
where $R=6\big(\dot{H}+2H^{2}\big)$ is the Ricci scalar.

Let us assume that both the homogeneous scalar field $\psi(t)$ and the 
potential $U(\psi)$ depend on time through the scale factor, i.e.
\begin{equation}
\psi(t)=\psi(a(t)),\quad U(\psi)=U(\psi(a));
\label{eq:5}
\end{equation}
then due to this simplified assumption the coefficient of the equation of state
$w_{\psi}$ is parameterized by the scale factor only
\begin{equation}
w_{\psi}=w_{\psi}(a),\quad p_{\psi}=w_{\psi}(a)\rho_{\psi}(a),
\label{eq:6}
\end{equation}
and
\begin{equation}
w_{\psi} = \frac{-\frac{1}{2}\psi'^{2}H^{2}a^{2} - U(\psi) + \xi \big[ 2
(\psi^{2})' H^{2}a + (\psi^{2})\ddot{}\big] + \xi \big[ \dot{H}
+3H^{2}\big]\psi^{2}}
{-\frac{1}{2}\psi'^{2}H^{2}a^{2} + U(\psi) - 3 \xi H^{2} \psi^{2} - 3 \xi
(\psi^{2})' H^{2}a}
\label{eq:7}
\end{equation}
where prime denotes the differentiation with respect to the scale factor.

We assume the flat model with the FRW geometry, i.e., the line element has the 
form
\begin{equation}
\ud s^{2} = -\ud t^{2} + a^{2}(t)[\ud r^{2} + r^{2}(\ud\theta^{2}
+ \sin^{2}{\theta}\ud\varphi^{2})],
\label{eq:8} 
\end{equation}
where $0 \leq \varphi \leq 2\pi$, $0 \leq \theta \leq \pi$ and $0 \leq r \leq
\infty$ are comoving coordinates, $t$ stands for the cosmological time. It is 
also assumed that a source of gravity is the phantom scalar field $\psi$ with 
the conformal coupling to gravity $\xi=1/6$. The dynamics is governed by the 
action
\begin{equation}
S=\frac{1}{2}\int \ud^{4}x \sqrt{-g}\left[m_{p}^{2}R +
(g^{\mu\nu}\psi_{\mu}\psi_{\nu} + \frac{1}{6} R\psi^{2} - 2 U(\psi))\right]
\label{eq:9}
\end{equation}
where $m_{p}^{2}=(8\pi G)^{-1}$; for simplicity and without lost of generality
we assume $4\pi G/3=1$ and $U(\psi)$ is the scalar field potential
\begin{equation}
U(\psi) = \frac{1}{2}m^{2}\psi^{2}.
\label{eq:10}
\end{equation}
After dropping the full derivatives with respect to time, rescaling phantom
field $\psi \rightarrow \phi = \psi a$ and the time variable to the conformal 
time $ \ud t = a \ud\eta$ we obtain the energy conservation condition
\begin{equation}
\mathcal{E} = \frac{1}{2}a'^{2} + \frac{1}{2}\phi'^{2} -
\frac{1}{2}m^{2}a^{2}\phi^{2}=\rho_{r,0}
\label{eq:11}
\end{equation}
where $\rho_{r,0}$ is constant corresponding to the radiation in the model. 
The equations of motion are
\begin{equation}
\left\{ \begin{array}{l}
a'' = m^{2}a\phi^{2},\\
\phi'' = m^{2} a^{2} \phi
	\end{array} \right.
\label{eq:12}
\end{equation}
where a prime denotes the differentiation with respect to the conformal time 
$\ud t = a \ud \eta$ and $m^{2}>0$.

From the energy conservation condition we have
\begin{equation}
\frac{1}{2}a'^{2}=\frac{\rho_{r,0}+\frac{1}{2}m^{2}a^{2}\phi^{2}}
{1+\dot{\phi}^{2}}
\label{eq:15}
\end{equation}
and now from the equations of motion (\ref{eq:12}) we receive
\begin{equation}
(\rho_{r,0}+\frac{1}{2}m^{2}a^{2}\phi^{2})\ddot{\phi} +
\frac{1}{2}m^{2}a\phi(1+\dot{\phi}^{2})(\phi\dot{\phi}-a)=0.
\label{eq:16}
\end{equation}

The effective equation of state parameter is 
\begin{equation}
w_{\text{eff}} = \frac{p_{\phi}+\frac{1}{3}\rho_{r}}{\rho_{\phi}+\rho_{r}},
\label{eq:17}
\end{equation}
for our model this parameter reduces to
\begin{equation}
w_{\text{eff}} = -\frac{1}{3}\frac{\frac{1}{2}\phi'^{2} +
\frac{1}{2}m^{2}a^{2}\phi^{2} - \rho_{r,0}}{-\frac{1}{2}\phi'^{2} +
\frac{1}{2}m^{2}a^{2}\phi^{2} + \rho_{r,0}}
\label{eq:18}
\end{equation}
where a prime denotes the differentiation with respect to the conformal time 
and finally taking into account equation~(\ref{eq:15}) we have
\begin{equation}
w_{\text{eff}} = -\frac{1}{3} \left[ \dot{\phi}^{2}
+ \frac{\frac{1}{2}m^{2}a^{2}\phi^{2} - \rho_{r,0}}
{\frac{1}{2}m^{2}a^{2}\phi^{2} + \rho_{r,0}}(1+\dot{\phi}^{2}) \right].
\label{eq:19}
\end{equation}

For $a,\phi \gg \rho_{r,0}$ this equation reduces to
\begin{equation}
w_{\text{eff}}=-\frac{1}{3}(2\dot{\phi}^{2}+1),
\label{eq:20}
\end{equation}
and it is clear that for any value of $\dot{\phi}$ $w_{\text{eff}}$ is always
negative.

To analyze equation (\ref{eq:16}) we reintroduce the original phantom field 
variable $\psi=\frac{\phi}{a}$ and $\ud a /a= \ud\ln{a}$. Now equation 
(\ref{eq:16}) reads
\begin{equation}
\bigg(\frac{\rho_{r,0}}{a^{4}} + \frac{1}{2}m^{2}\psi^{2}\bigg)(\psi'' + \psi') +
\frac{1}{2}m^{2}\psi(1+(\psi'+\psi)^{2})(\psi(\psi'+\psi)-1)=0
\label{eq:21}
\end{equation}
where a prime now denotes the differentiation with respect to a natural 
logarithm of the scale factor. Introducing new variables $y=\psi'$ and 
$\rho_{r}=\rho_{r,0}a^{-4}$ we can represent this equation as an autonomous 
dynamical system
\begin{align}
\psi' &= y \nonumber \\
y' &= -y-\frac{\frac{1}{2}m^{2}\psi}{\rho_{r}+\frac{1}{2}m^{2}\psi^{2}}
(\psi(y+\psi)-1)(1+(y+\psi)^{2}) \label{eq:22} \\
\rho_{r}' &= -4\rho_{r}. \nonumber
\end{align}

There are the two critical points in the phase space $(\psi,y,\rho_{r})$, namely
$\psi=\pm 1$, $y=0$, $\rho_{r}=0$. The linearization matrix reads
\begin{equation}
A = \left[ \begin{array}{ccc}
0 & 1 & 0\\
-2(1+\psi^{2}) & -1-(1+\psi^{2}) & \frac{2}{m^{2}\psi^{3}}(\psi^{2}-1)(1+\psi^{2})\\
0 & 0 & -4\\
        \end{array} \right]_{y=0,\rho_{r}=0} =
\left[ \begin{array}{rrr}
0 & 1 & 0\\
-4 & -3 & 0\\
0 & 0 & -4\\
	\end{array} \right]_{y=0,\rho_{r}=0,\psi=\pm 1} .
\label{eq:23}
\end{equation}
The eigenvalues for this matrix are $\lambda_{1,2}=\frac{1}{2}(-3\pm i \sqrt{7})$
and $\lambda_{3}=-4$.

To find a global phase portrait it is necessary to study the system in the
neighbourhood of the critical points which correspond, from the physical point 
of view, stationary states (or asymptotic solutions). Then the Hartman-Grobman 
theorem guaranties us that the linearized system at this point is a well 
approximation of the nonlinear system. First, we must note that $\rho_{r}=0$ is 
in the invariant submanifold of the $3$-dimensional nonlinear system. It is 
also useful to calculate the eigenvectors for any eigenvalue. We obtain 
following eigenvectors
\begin{equation}
v_{1,2}=\left[ \begin{array}{r}
-\frac{3}{8} \\ 1 \\ 0 \\
		\end{array} \right] \mp
	i\left[ \begin{array}{c}
\frac{\sqrt{7}}{8} \\ 0 \\ 0 \\
		\end{array} \right], \qquad
v_{3}=\left[ \begin{array}{c}
0 \\ 0 \\ 1 \\
	     \end{array} \right].	
\end{equation}
They are helpful in construction of the exact solution of the linearized system
\begin{equation}
\vec{x}(t)=\vec{x}(0)\exp{t \left[ \begin{array}{rrr}
0 & 1 & 0 \\
-4 & -3 & 0 \\
0 & 0 & -4 \\
\end{array} \right]} = 
\left[ \begin{array}{rrr}
0 & -\frac{3}{8} & \frac{\sqrt{7}}{8} \\
0 & 1 & 0 \\
1 & 0 & 0 \\
\end{array} \right]
\left[ \begin{array}{rrr}
e^{-4t} & 0 & 0 \\
0 & e^{-\frac{3}{2}t}\cos{\frac{\sqrt{7}}{2}t} &
-e^{-\frac{3}{2}t}\sin{\frac{\sqrt{7}}{2}t} \\
0 & e^{-\frac{3}{2}t}\sin{\frac{\sqrt{7}}{2}t} &
e^{-\frac{3}{2}t}\cos{\frac{\sqrt{7}}{2}t} \\
\end{array} \right]
\left[ \begin{array}{rrr}
0 & 0 & 1 \\
0 & 1 & 0 \\
\frac{8}{\sqrt{7}} & \frac{3}{\sqrt{7}} & 0 \\
\end{array} \right] 
\left[ \begin{array}{c}
 x_{0} \\ y_{0} \\ z_{0} \\
 \end{array} \right] .
\end{equation}
where $x=\psi - \psi_{0}$, $y=\psi' - \psi_{0}'$, $z=\rho_{r} - \rho_{0}$ and
$x_{0}, y_{0}, z_{0}$ are initial conditions and we have substituted $\ln{a}=t$.

If we consider linearized system on the invariant stable submanifold $z=0$, it
is easy to find the exact solution. If we return to the original variables $\psi,
\psi'$, then $\psi(\ln{a})$ is the solution of the linear equation
\begin{equation}
(\psi - \psi_{0})'' + 3(\psi - \psi_{0})' + 4(\psi-\psi_{0})=0,
\end{equation}
i. e.,
\begin{equation}
(\psi-\psi_{0})=C_{1}\exp{\left(-\frac{3}{2}\ln{a}\right)} 
\cos{\left(\frac{\sqrt{7}}{2}\ln{a}\right)} +
C_{2}\exp{\left(-\frac{3}{2}\ln{a}\right)} 
\sin{\left(\frac{\sqrt{7}}{2}\ln{a}\right)}
\end{equation}
or
\begin{equation}
(\phi-\phi_{0})=C_{1}a^{-\frac{1}{2}}\cos{\left(\frac{\sqrt{7}}{2}\ln{a}\right)} +
C_{2}a^{-\frac{1}{2}}\sin{\left(\frac{\sqrt{7}}{2}\ln{a}\right)}.
\end{equation}

\begin{figure}
\includegraphics[scale=0.5]{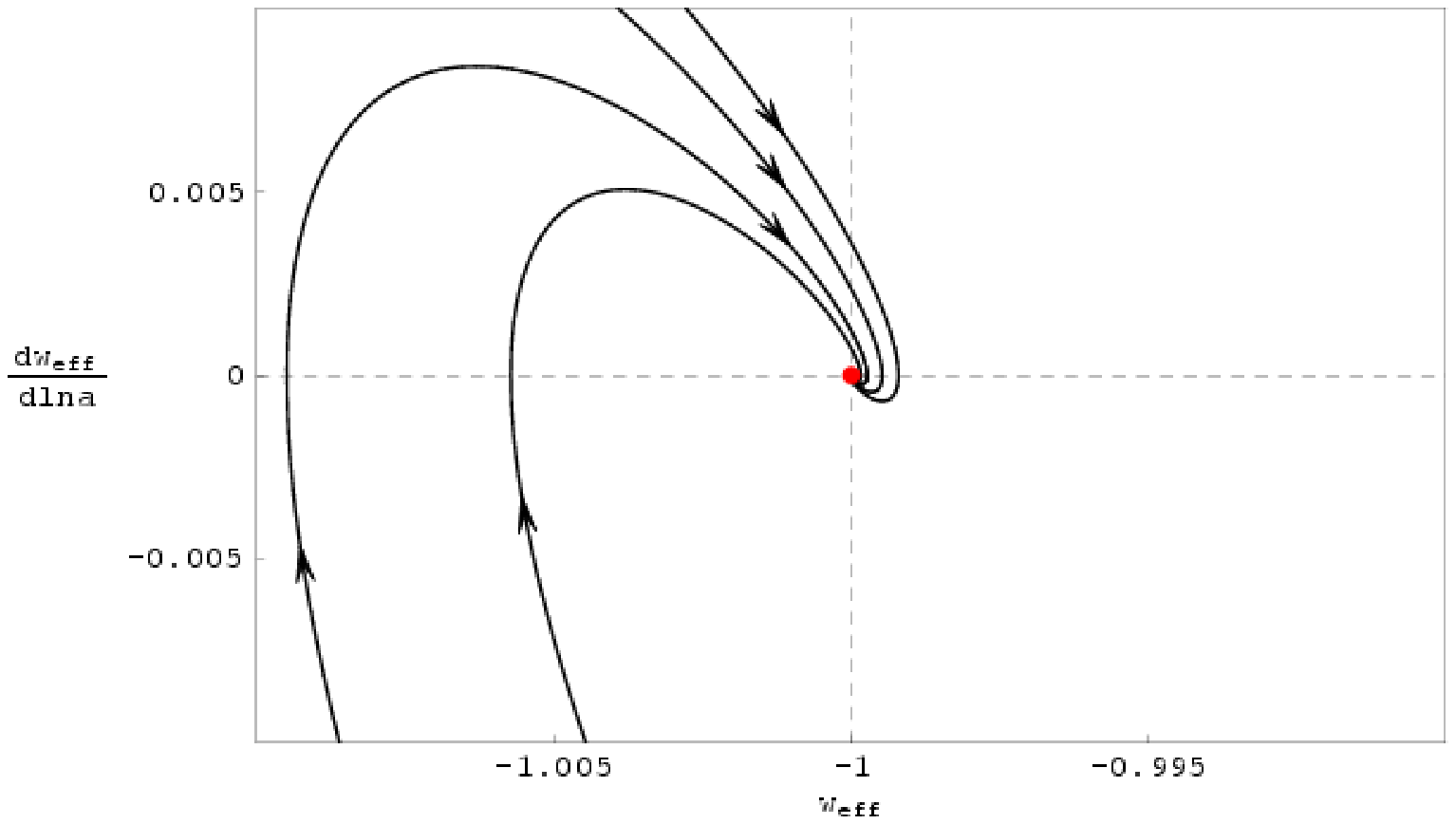}
\includegraphics[scale=0.5]{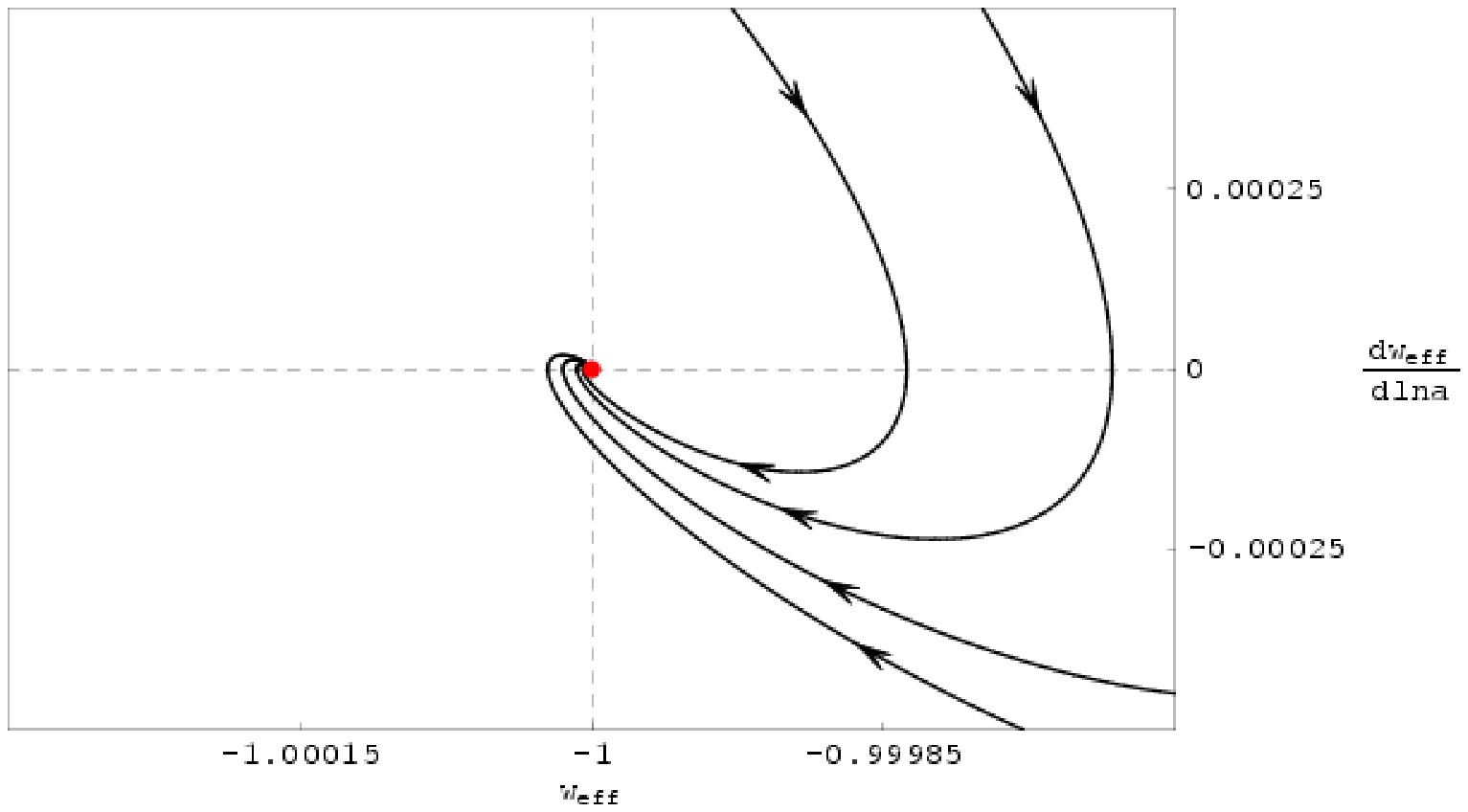}
\caption{The phase portrait $(w_{\text{eff}},w_{\text{eff}}')$ of the 
investigated model on the submanifold $\rho_{r}=0$. This figure 
illustrates the evolution of the dark energy equation of the state parameter 
as a function of redshift for different initial conditions. In all cases 
trajectories cross the boundary line $w_{\text{eff}}=-1$ infinite many times 
but this state also represents the global attractor.}
\label{fig:1}
\end{figure}

\begin{figure}
\includegraphics[scale=1]{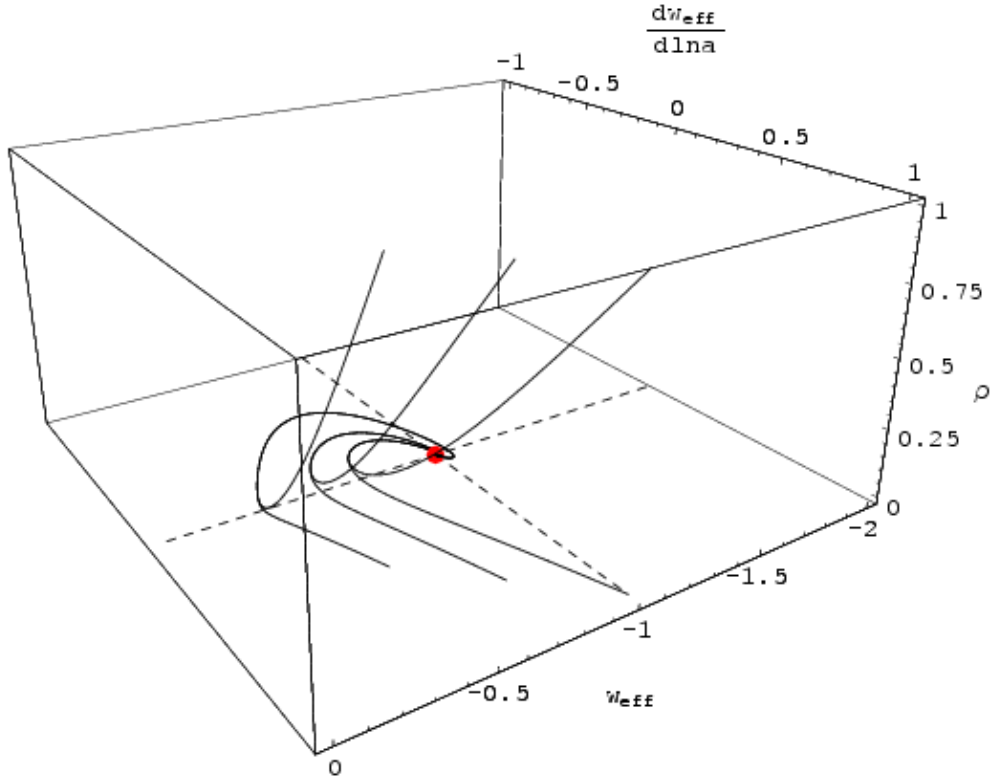}
\caption{The phase portrait of the $3$-dimensional dynamical system 
(\ref{eq:22}) in terms of the variables 
$(w_{\text{eff}},w'_{\text{eff}},\rho_{r})$ and projections of trajectories on 
the submanifold $(w_{\text{eff}},w'_{\text{eff}},0)$. The critical 
point represents the state where $w_{\text{eff}}=-1$ -- the cosmological 
constant. The trajectories approach this point as the scale factor goes to 
infinity (or the redshift $z\to-1$). Before this stage the weak energy 
condition is violated infinite number of times.}
\label{fig:2}
\end{figure}

Because of the lack of alternatives to the mysterious cosmological constant
\cite{Kurek:2007,Szydlowski:2006kk} we allow that energy might vary in time
following assumed a priori parameterization of $w(z)$. In the popular 
parameterization \cite{Gerke:2002,Linder:2003,Cooray:1999,Padmanabhan:2003} 
appears a free function in most scenarios which is a source of difficulties in 
constraining parameters by observations. However most parameterizations of the 
dark energy equation of state cannot reflect real dynamics of cosmological  
models with dark energy. The assumed form of $w(z)$ can be incompatible with 
the $w(z)$ obtained from the underlying dynamics of the cosmological model. 
For example some of parameters can be determined from the dynamics which 
can be crucial in testing and selection of cosmological models \cite{Kurek:2007}. 
Our point of view is to obtain the form of $w(z)$ specific for given class of 
cosmological models from dynamics of this models and apply it in further 
analysis both theoretical and empirical. In practice we put the cosmological 
model in the form of the dynamical system and linearize it around the 
neighbourhood of the present epoch to find the exact formula of $w(z)$. 
For the phantom scalar field model this incompability manifests by the presence 
of a focus type critical point (therefore damping oscillations) in the phase 
space rather than a stable node (Fig.~\ref{fig:1} and its $3D$ version
Fig.~\ref{fig:2}).

The properties of the minimally coupled phantom field in the FRW cosmology 
using the phase portrait have been investigated by Singh et al. 
\cite{Singh:2003} (see also \cite{Wei:2006} for more recent studies). Authors
showed the existence of the deSitter attractor and damped oscillations (the 
ratio of the kinetic to the potential energy $|T/U|$ to oscillate to zero).

We can also express $w_{\text{eff}}$ in these new variables
\begin{equation}
w_{\text{eff}} = -\frac{1}{3} \bigg\{ (\psi+\psi')^{2} -
\frac{\rho_{r}-\frac{1}{2}m^{2}\psi^{2}}{\rho_{r} + \frac{1}{2}m^{2}\psi^{2}}
\left[ 1+(\psi+\psi')^{2} \right] \bigg\}
\label{eq:24}
\end{equation}
\begin{equation}
w_{\text{eff}}'=\frac{\ud w_{\text{eff}}}{\ud \ln{a}}=-\frac{2}{3} \bigg\{ 
\frac{m^{2}\psi^{2}}{\rho_{r}+\frac{1}{2}m^{2}\psi^{2}}(\psi+\psi')(\psi'+\psi'') 
+ \rho_{r}m^{2}\psi^{2}\frac{2+\frac{\psi'}{\psi}}{(\rho_{r} 
+\frac{1}{2}m^{2}\psi^{2})^{2}} \left[ 1+(\psi+\psi')^{2} \right]\bigg\}.
\label{eq:25}
\end{equation}

Recently Caldwell and Linder \cite{Caldwell:2005} have discussed dynamics of
quintessence models of dark energy in terms of $w-w'$ phase variables, where
$w'$ was the differentiation with respect to the logarithm of the scale factor. 
These methods were extended to the phantom and quintom models of dark energy 
\cite{Chiba:2006, Guo:2006}. Guo et al. \cite{Guo:2006} examined the two-field 
quintom models as the illustration of the simplest model of transition across 
the $w_{X}=-1$ barrier. The interesting mechanism of acceleration with a 
periodic crossing of the $w=-1$ barrier have been recently discussed in the 
context of the cubic superstring field theory \cite{Aref'eva:2006et}. In the 
model under consideration we obtain this effect but trajectories cross the 
barrier infinitely many times. The main advantage of the discovered road to 
$\Lambda$ is that it takes place in the simple flat FRW model with the quadratic 
potential of the scalar field.

It is easy to check that at the critical points $w_{\text{eff}}=-1$ and 
$\frac{\ud w_{\text{eff}}}{\ud \ln{a}}=0$. Since these 
points are sinks there is infinite many crossings of $w_{\text{eff}}=-1$ during 
the evolution.

The methods of the Lyapunov function are useful in discussion of stability of
the critical point of the non-linear system. The stability of any hyperbolic
critical point of dynamical system is determined by the signs of the real parts
of the eigenvalues $\lambda_{i}$ of the Jacobi matrix. A hyperbolic critical
point is asymptotically stable iff real $\lambda_{i}<0$ $\forall i$, if
$x_{0}$ is a sink. The hyperbolic critical point is unstable iff it is either a
source or a saddle. The method of the Lyapunov function is especially useful in
deciding the stability of a non-hyperbolic critical points \cite[p.129]{Perko:1991}.
The construction of the Lyapunov function was used by \cite{Giacomini:2006ak} for
demonstration that periodic behaviour of a single scalar field is not possible
for minimally coupled phantom scalar field (see also \cite{Castagnino:1999wd}).

\begin{figure}
\includegraphics[scale=1]{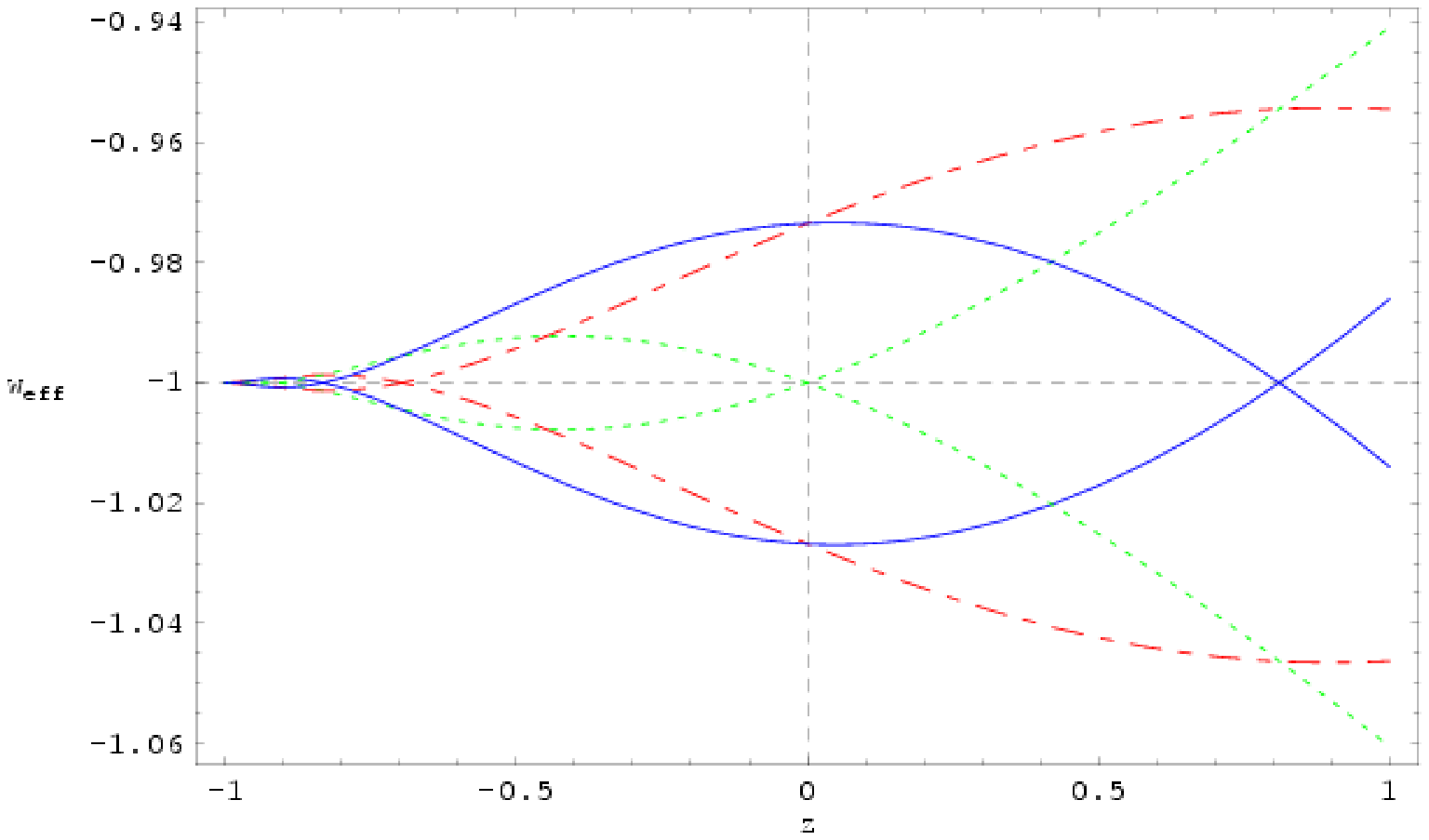}
\caption{The relation $w(z)$ for different values of the parameters:
$C_{1}=0.01$, $C_{2}=0$ dash-dotted line (red el. version); $C_{1}=0$,
$C_{2}=-0.01$ dotted line (green el. version); $C_{1}=0.01$, $C_{2}=-0.01$
solid line (blue el. version) and $\psi_{0}=\pm 1$. For all solutions $w(z)$
approaches the cosmological constant as $z\to -1$. It is obvious that this 
relation is true only in the vicinity of the critical point
$(w_{\text{eff}},w'_{\text{eff}},\rho_{r})=(-1,0,0)$.}
\label{fig:3}
\end{figure}

The quantity $w_{\text{eff}}'$ in terms of $w_{\text{eff}}$ and $(\ln{\psi})'$ reads
\begin{equation}
w_{\text{eff}}'=-(1-3w_{\text{eff}})(1+w_{\text{eff}}+\frac{2}{3}(\ln{\psi})').
\label{eq:26}
\end{equation}
It is interesting that equation~(\ref{eq:26}) can be solved in terms of
$\bar{w}(a)$ -- the mean of the equation of the state parameter in the 
logarithmic scale defined by Rahvar and Movahed \cite{Rahvar:2007} as
\begin{equation}
\bar{w}(a)=\frac{\int_{1}^{a}w(a')\ud(\ln{a'})}{\int_{1}^{a}\ud(\ln{a'})},
\end{equation}
namely:
\begin{equation}
w(a)=\frac{1}{3}-\frac{4}{3}a^{3(1+\bar{w}(a))}\psi^{2}.
\end{equation}
They argued that this phenomenological parameterization removes the fine tuning 
of dark energy and $\rho_{X}/\rho_{m} \propto a^{-3\bar{w}(a)}$ approaches 
a unity at the early universe. Note that in $\bar{w}(a)=-1$ that
\begin{equation}
w(z)+1=\frac{4}{3}(1-\psi^{2}),
\end{equation}
where 
\begin{equation}
\psi=\psi_{0} + (1+z)^{\frac{3}{2}}\bigg\{C_{1}\cos(\frac{\sqrt{7}}{2}\ln(1+z)) -
C_{2}\sin(\frac{\sqrt{7}}{2}\ln(1+z))\bigg\}.
\end{equation}
In Fig.~\ref{fig:3} we present the relation $w(z)$ for different values of 
parameters $\psi_{0}=\pm 1$, $C_{1}$ and $C_{2}$.

In this letter we regarded the phantom scalar field conformally coupled to 
gravity in the context of the problem of acceleration of the Universe. We 
applied the methods of dynamical systems and the Hartman-Grobman theorem 
to find universal behaviour at the late times -- damping oscillations around 
$w_{\text{eff}}=-1$. We argued that most parameterizations of the dark energy, 
such as linear evolution of $w(z)$ in redshift or the scale factor, cannot 
reflect realistic physical models because of the presence of non-hyperbolic 
critical point of a focus type on the phase plane $(w,w')$. We suggested a 
parameterization of a type 
\begin{equation}
w_{X}(z)=-1+(1+z)^{3}\bigg\{C_{1}\cos(\ln(1+z)) + C_{2}\sin(\ln(1+z))\bigg\}
\end{equation}
which parameterizes damping oscillations around $w_{X}=-1$ ``phantom divide'',
and finally, with the help of this formula one can simply calculate energy
density for dark energy $\rho_{X}$
\begin{equation}
\rho_{X}=\rho_{X,0}\exp{\big(-B\big)}
\exp{\big((1+z)^{3}\big[A\sin(\ln(1+z))+B\cos(\ln(1+z))\big]\big)}.
\end{equation}

\begin{acknowledgments}
The work of M.S. has been supported by the Marie Curie Actions
Transfer of Knowledge project COCOS (contract MTKD-CT-2004-517186).
\end{acknowledgments}

\end{document}